\documentclass[fleqn,10pt]{wlscirep}
\usepackage[utf8]{inputenc}
\usepackage[T1]{fontenc}
\usepackage{lineno}
\usepackage{multirow}
\usepackage{bm}
\usepackage{comment}
\usepackage{subfigure}
\usepackage{makecell}
\usepackage{url}
\usepackage{hyperref} 
\title{SynthSoM: A synthetic intelligent multi-modal sensing-communication dataset for Synesthesia of Machines (SoM)}
\author[1,*]{Xiang Cheng$\dag$}
\author[1]{Ziwei Huang$\dag$}
\author[1]{Yong Yu$\dag$}
\author[2,3,4*]{Lu Bai}
\author[1]{Mingran Sun}
\author[1]{Zengrui Han}
\author[1]{Ruide Zhang}
\author[1]{Sijiang Li}
\affil[1]{State Key Laboratory of Photonics and Communications, School of Electronics, Peking University, Beijing, 100871, China}
\affil[2]{Joint SDU-NTU Centre for Artificial Intelligence Research (C-FAIR), Shandong University, Jinan, 250100, China}
\affil[3]{National Mobile Communications Research Laboratory, Southeast University, Nanjing, 214135, China}
\affil[4]{Shandong Research Institute of Industrial Technology, Jinan, 250100, China}
\affil[*]{corresponding authors: Xiang Cheng (xiangcheng@pku.edu.cn); Lu Bai (lubai@sdu.edu.cn)}
\affil[$\dag$]{these authors contributed equally to this work}

\begin{abstract}
Given the importance of datasets for sensing-communication integration research,  a novel simulation platform for constructing  communication and multi-modal sensory dataset is developed. The developed platform integrates three high-precision software, i.e., AirSim, WaveFarer, and Wireless InSite, and further achieves in-depth integration and precise alignment of them. Based on the developed platform, a new synthetic intelligent multi-modal sensing-communication dataset for Synesthesia of Machines (SoM), named SynthSoM, is proposed. The SynthSoM dataset contains various air-ground multi-link cooperative scenarios with comprehensive conditions, including multiple weather conditions, times of the day, intelligent agent densities, frequency bands, and antenna types. The SynthSoM dataset encompasses multiple data modalities, including radio-frequency (RF) channel large-scale and small-scale fading data, RF millimeter wave (mmWave) radar sensory data, and non-RF sensory data, e.g., RGB  images, depth maps, and light detection and ranging (LiDAR) point clouds.  The quality of SynthSoM dataset is validated via statistics-based qualitative inspection and evaluation metrics through machine learning (ML) via real-world measurements. The SynthSoM dataset is open-sourced and provides consistent data for  cross-comparing SoM-related algorithms. 

\end{abstract}
\begin{document}

\flushbottom
\maketitle

\thispagestyle{empty}

\section*{Background \& Summary}
In sixth-generation (6G) enabled artificial intelligence of things (AIoT), intelligent agents will be equipped with multi-modal sensors and communication devices \cite{cheng2023intelligent, chen2022computer,mundlamuri2023sensing}. By processing the collected multi-modal sensory and communication data, intelligent agents can complete multi-modal sensing-communication tasks \cite{steyer2019grid,charan2021vision}. However, since integrated sensing and communications (ISAC) focuses on the integration of radio-frequency (RF) communications and sensing \cite{he2023integrated,liu2022integrated,jornet2023wireless}, it cannot support multi-modal sensing-communication tasks. To fill this gap, a novel concept, named Synesthesia of Machines (SoM), is proposed \cite{cheng2023intelligent}. Unlike ISAC, SoM refers to intelligent multi-modal sensing-communication integration, containing  communications, RF sensing, i.e., millimeter wave (mmWave) radar, and non-RF sensing, i.e., light detection and ranging (LiDAR) and RGB-D cameras. To achieve mutual enhancement among communications and multi-modal sensing, artificial neural network (ANN) serves as a fundamental tool owing to its advantage in exploring nonlinear relationships \cite{chen2023follownet,jeon2024purely,aguirre2024hardware}.
As the cornerstone of SoM research via ANN, a comprehensive dataset is important. Currently, extensive datasets have been constructed \cite{li2022high,sun2022shift,xu2022opv2v,qi2022open,manivasagam2020lidarsim,geiger2013vision,alkhateeb2023deepsense,li2024synthetic,alrabeiah2020viwi}.  
Several typical datasets are described below. 

\textbf{KITTI}: The KITTI dataset is a multi-modal measurement dataset \cite{geiger2013vision}, while
the communication and mmWave radar sensory information were ignored.

\textbf{DeepSense 6G}:
A multi-modal measurement dataset, named DeepSense 6G,  was constructed \cite{alkhateeb2023deepsense}. However, depth maps and a weather condition, i.e., snowy day, were ignored. Although the measurement dataset \cite{alkhateeb2023deepsense}  facilitates method validation, it is difficult to customize multi-modal sensing-communication scenarios owing to cost concerns.

\textbf{SDCD}: By using efficient software, a multi-modal synthetic dataset, named Synthetic Digital City Dataset (SDCD), was constructed \cite{li2024synthetic}. Nevertheless, the SDCD lacked communication, mmWave radar, and LiDAR information. 

\textbf{ViWi}: The Vision-Wireless (ViWi) dataset is a synthetic multi-modal sensing-communication  dataset, including channel information, RGB images, depth maps, and LiDAR point clouds \cite{alrabeiah2020viwi}. However, UAV scenarios, mmWave radar information,  times of the day, and  weather conditions were ignored. 

Although the existing dataset plays an essential role in SoM research, it has some limitations. The KITTI, DeepSense 6G, SDCD, and ViWi datasets lacked comprehensive communication and multi-modal sensory data. Furthermore, for multi-modal sensing-communication datasets, i.e., DeepSense 6G and ViWi, the former requires expensive measurement devices with limited flexibility, while the latter has simple simulation scenarios with low rendering effect. 

Our previous work \cite{cheng2023m} attempted to fill the aforementioned gap and constructed a synthetic multi-modal sensing-communication dataset, named M$^3$SC \cite{cheng2023m}. However, the M$^3$SC dataset \cite{cheng2023m} lacked channel large-scale fading data. Furthermore, the M$^3$SC dataset focused on a typical scenario, i.e., vehicular urban crossroad, and a certain intelligent agent destiny. The M$^3$SC dataset also lacked technical validation based on real-world measurements and the code was not open-sourced. Therefore, an open-sourced dataset, which encompasses channel large/small-scale fading information and multi-modal sensory data, incorporates the interaction of various air-ground multi-link cooperative scenarios and conditions, and is validated by real-world measurements, is still lacking. 

A new open-sourced synthetic intelligent multi-modal sensing-communication dataset for SoM, named SynthSoM, is constructed, as shown in Fig.~\ref{fig:step}. The comparison between the SynthSoM dataset and the aforementioned datasets \cite{alrabeiah2020viwi,li2024synthetic,alkhateeb2023deepsense,geiger2013vision,cheng2023m} is  given in Table~\ref{tab:example1}.   We summarize the main contributions and novelties of this work:

\begin{table}[!t]
\centering
\begin{footnotesize}
\begin{tabular}{|l|l|l|l|l|l|l|l|l|l|l|l|}
\hline
\multirow{2}{*}{\textbf{Dataset}}& \multirow{2}{*}{\textbf{Comm.}} & \multicolumn{4}{|l|}{\textbf{Sensing}} & \multirow{2}{*}{\textbf{\makecell[l]{Air-Ground Multi-Link \\Cooperative Scenario}}} & \multicolumn{3}{|l|}{\textbf{Weather}} & \multirow{2}{*}{\textbf{Time}}  & \multirow{2}{*}{\textbf{Type}}\\
\cline{3-6} 
\cline{8-10} 
& & Radar & RGB & Depth Map & LiDAR & & Sunny & Rainy & Snowy & &    \\
\hline
\textbf{KITTI}\cite{geiger2013vision} & \bm{$\times$} & \bm{$\times$} & \checkmark & \checkmark & \checkmark  &  \bm{$\times$}  & \checkmark  & \bm{$\times$}  & \bm{$\times$}   & \bm{$\times$}  & Meas.\\ 
\hline
\textbf{DeepSense 6G}\cite{alkhateeb2023deepsense} & Limited & \checkmark & \checkmark & \bm{$\times$} & \checkmark  & \bm{$\times$}   & \checkmark  & \checkmark  & \bm{$\times$}   & \checkmark & Meas.  \\ 
\hline
\textbf{SDCD}\cite{li2024synthetic} & \bm{$\times$} & \bm{$\times$} & \checkmark & \checkmark  & \bm{$\times$}   & \bm{$\times$}  & \checkmark  & \checkmark  & \checkmark    & \checkmark & Synth.  \\
\hline
\textbf{ViWi}\cite{alrabeiah2020viwi} & Limited & \bm{$\times$} & \checkmark & \checkmark  & \checkmark   & \bm{$\times$} & \checkmark  & \bm{$\times$}  & \bm{$\times$}    & \bm{$\times$} & Synth.   \\
\hline
\textbf{M$^3$SC}\cite{cheng2023m} & Limited & \checkmark & \checkmark & \checkmark  & \checkmark  & \bm{$\times$} & \checkmark  & \checkmark  & \checkmark    & \checkmark & Synth.   \\
\hline
\textbf{SynthSoM} & \checkmark & \checkmark & \checkmark & \checkmark  & \checkmark     & \checkmark & \checkmark & \checkmark & \checkmark    & \checkmark & Synth.   \\
\hline
\end{tabular}
\caption{\label{tab:example1}Comparison of the SynthSoM dataset and the existing datasets.}
\end{footnotesize}
\end{table}

\begin{itemize}
\item To construct the SynthSoM dataset for  intelligent multi-modal sensing-communication integration, 
a novel simulation platform  is developed. The developed platform integrates three software, i.e., AirSim, WaveFarer, and Wireless InSite, and further achieves in-depth integration and precise alignment of them. 

\item The SynthSoM dataset covers diverse air-ground multi-link cooperative scenarios, e.g., vehicles and unmanned aerial vehicles (UAVs) under urban, suburban, and rural areas. Also, the SynthSoM dataset contains comprehensive conditions, e.g., multiple weather conditions, times of the day, intelligent agent densities, frequency bands, and antenna types. The simultaneous incorporation and interaction  of various scenarios and conditions enhance the comprehensiveness of the SynthSoM dataset.
\item The SynthSoM dataset encompasses multiple data modalities, including RF communications, i.e., 140K sets of channel matrices and 18K sets of path loss, RF sensing, i.e.,  136K sets of mmWave radar waveforms with 38K radar point clouds, and non-RF sensing, i.e., 145K RGB images, 290K depth maps, and 79K sets of LiDAR point clouds.
\item The quality of the SynthSoM dataset is validated in two ways, i.e., statistics-based qualitative inspection and ML-based evaluation metrics via real-world measurements. Our open-sourced SynthSoM dataset and codebase can provide the useful insight for SoM-related algorithm cross comparison, model calibration, and baseline implementation.
\end{itemize}


\section*{Methods}
\begin{figure}[!t]
\centering
\includegraphics[width=0.99\linewidth]{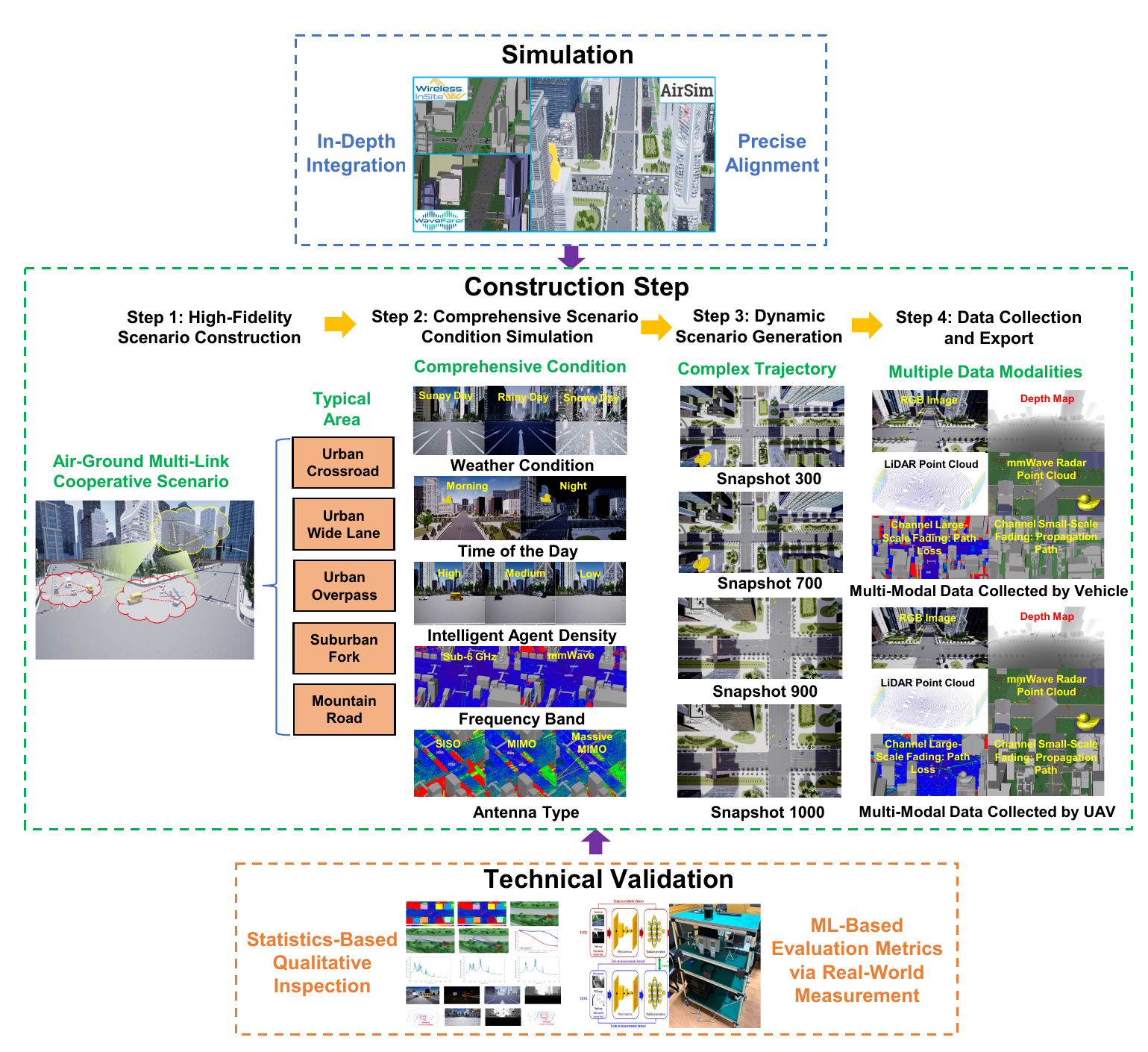}
\caption{Framework of constructing the developed simulation platform for the generation of the SynthSoM dataset.}
\label{fig:step}
\end{figure}

To generate and collect communication and multi-modal sensory data, the developed simulation platform utilizes AirSim, WaveFarer, and Wireless InSite, and further achieves in-depth integration and precise alignment of them. To achieve in-depth integration, different scenario and condition files are properly imported into AirSim, WaveFarer, and Wireless InSite. Furthermore, three-dimensional (3D) coordinates of intelligent agents are set frame by frame to imitate the continuous motion of multi-vehicles/UAVs in different software. To achieve precise alignment, scenarios and conditions in AirSim, WaveFarer, and Wireless InSite are  the same.
In addition, the size and trajectory of each object are set to be the same in each software. 
Overall, there are four steps involved in developing the simulation platform, as shown in Fig.~\ref{fig:step}.
The first step, i.e., high-fidelity scenario construction,  constructs the aligned sensing and communication scenarios in AirSim, WaveFarer, and Wireless InSite
at the initial snapshot. The second step, i.e., comprehensive scenario condition simulation, sets various conditions, such as weather conditions, times of the day, intelligent agent densities, frequency bands, and antenna types. The third step, i.e., dynamic scenario generation, generates the  sensing and communication scenarios with the aligned trajectories of dynamic intelligent agents in AirSim, WaveFarer, and Wireless InSite at other snapshots. The fourth step, i.e., data collection and export, aims to collect and export communication and multi-modal sensory data automatically. 

\subsection*{High-Fidelity Scenario Construction}
Procedures of high-fidelity scenario construction in the developed simulation platform with the help of AirSim, WaveFarer, and Wireless InSite  are presented as follows.

To generate and collect the non-RF sensory data, i.e., RGB images, depth maps, and LiDAR point clouds, we utilize an open-sourced software developed by Microsoft, i.e., AirSim \cite{shah2018airsim}. AirSim provides a virtual 3D environment to simulate diverse scenarios and conditions. First, by exploiting a city modeling software, i.e., CityEngine \cite{CityEngine}, we utilize  satellite maps in the real world  to generate building models, which are exported to Unreal Engine 4 (UE4) \cite{UE}. A powerful software, i.e., RoadRunner \cite{RoadRunner}, is further utilized to conduct detailed processing on road layout, terrain features, and plants, which are also exported to UE4. Furthermore, we place  intelligent agents, such as UAVs and vehicles (buses and cars), within the scenario. Finally, we refine the scenario details in UE4 and perform in-depth rendering. In the constructed scenario, the intelligent agent and roadside facility (RSF) are equipped with multi-modal sensors, including RGB cameras, depth cameras, and LiDAR devices. First, to collect RGB images, AirSim utilizes the rendering pipeline of UE4 to
simulate camera imaging. Each simulated camera is equipped with a SceneCaptureComponent2D
component from UE4, which captures the scenario from the camera frustum and renders the view. Second, to collect depth maps, UE4 generates a depth buffer of the scenario for each snapshot during rendering. Through the SceneCapture component, the depth value corresponding to each pixel within the
camera frustum can be captured. Third, to collect LiDAR point clouds, AirSim simulates a LiDAR by emitting
a multitude of laser rays into space and detecting the intersection points of these rays with objects to
obtain distance points. In addition, AirSim captures the occlusion, reflections, and sensor noise/failure by utilizing Z-Buffer method to achieve pixel-level occlusion effects, exploiting physically-based reflection properties, and adding noise models to sensors to simulate real-world measurement inaccuracies, respectively.
According to the requirement of different tasks, key parameters of multi-modal sensors, such as field-of-view (FoV), resolution, LiDAR range, and the number of LiDAR channels, can be adjusted. 

To generate and collect the RF sensory data, i.e., mmWave radar information, we utilize efficient commercial software developed by REMCOM, i.e., WaveFarer \cite{WaveFarer}. WaveFarer exploits the advanced computational approach, such as full-wave electromagnetic simulation, to characterize wave propagation. First, we import 3D scenario models from the project of AirSim into WaveFarer to achieve the in-depth integration  between AirSim and WaveFarer. Specifically, we establish the units for 3D scenario model importation to ensure the accuracy of the scenario dimensions. Subsequently, we import the STereoLithography (STL) format models provided by AirSim and assign the corresponding material properties, such as wood, metal, and concrete, according to the object categories, such as trees, vehicles, and buildings.
Second, we set materials for 3D scenario models and configure the electrical conductivity and magnetic permeability of  materials. Third, we configure the diffuse scattering parameter to ensure that the mmWave radar can detect objects. Finally, we import intelligent agent models into WaveFarer and set their electromagnetic materials. In the constructed scenario, intelligent agents and RSF are equipped with mmWave radars, which leverage the linear frequency modulated continuous wave (FMCW) operating in the frequency range of $77$ GHz to $81$ GHz. The sampling interval for mmWave radar information is $5$ nanoseconds, with each chirp containing $1000$ sampling points and each snapshot consisting of $101$ chirps, resulting in a waveform file length of $101,000$ samples per snapshot.
Furthermore, to collect mmWave radar information, the ``Create Linear Chirp Simulation'' script provided by WaveFarer
is utilized to perform ray-tracing simulation for FMCW radar and to acquire the echo signals. Then, the official ``Generate Range-Doppler'' script is employed to process the echo
signals obtained from the previous step, simulating the down-conversion and filtering operations of an actual
FMCW radar. Next, a three-dimensional fast Fourier transform (3D-FFT) is exploited to the processed echo
signals across the range, velocity, and angle dimensions, followed by the implementation of constant false
alarm rate (CFAR) detection to obtain the mmWave radar point cloud data. Additionally, WaveFarer captures the occlusion, reflections, and sensor noise/failure by importing 3D models, adjusting electromagnetic parameters
of object materials, and adjusting the detection threshold of the CFAR algorithm to characterize
different noise levels at the code level, respectively. Key parameters of mmWave radars, e.g., detection range, angular resolution, and sweep frequency range of FMCW, can be adjusted.

To generate and collect communication information, we utilize efficient commercial software developed by REMCOM, i.e., Wireless InSite \cite{WI}. Based on the powerful ray-tracing technology, Wireless InSite can imitate realistic wireless propagation effects, where RF interference and multipath propagation effects can be captured. For the capturing of RF interference, the ray-tracing technology simulates propagation paths of electromagnetic waves  by integrating reflection, refraction, diffraction, and scattering models. The utilization of ray-tracing technology can facilitate comprehensively analyzing the source  of RF interference and examining the interactions between different signal sources, thus evaluating the effect of RF interference. For the capturing of multipath propagation effects, a set of rays can be generated from the transmitter (Tx) with a specific angular separation. Then, rays are propagated in the environment up to a maximum number of interactions with surfaces, and are considered to reach the receiver (Rx) if rays meet receiving conditions.
First, we  import  models in AirSim into Wireless InSite to achieve the in-depth integration between AirSim and Wireless InSite. Similar to the previously mentioned in-depth integration with AirSim and WaveFarer, the STL format 3D scenario models provided by AirSim are imported and the corresponding material settings, e.g., wood, metal, and concrete, are applied according to the object categories, e.g., trees, vehicles, buildings. Then, we set materials of different types of models. Moreover, we conduct the double size operation on imported models to enhance the completeness of the triangle mesh. Finally, we place the intelligent agent within the scenario and set the frequency band, antenna type, and waveform type. In the constructed scenario, intelligent agents and RSF are equipped with communication equipment. Key parameters of communication equipment, such as carrier frequency, bandwidth, waveform type, and antenna type, can also be adjusted. By utilizing the ray-tracing technology in different environments with different 3D scenario models, the RF interference and multipath propagation effects across different environments can be properly captured. Also, Wireless InSite captures the occlusion, reflections, and sensor noise/failure by simulating the blocking of signals via obstacles, utilizing shooting and bouncing ray and the uniform theory of diffraction methods, and providing hardware noise modeling functions, respectively.

Note that a detailed parameter setting in the first step, i.e., high-fidelity scenario construction, can be observed in \url{https://github.com/ZiweiHuang96/SynthSoM}. For example, for Scenario 1, i.e., Air-Ground Multi-Link Cooperative Urban Crossroad, key parameters, including sensor position, sensor parameter, and RSF  position, can be found in \url{https://github.com/ZiweiHuang96/SynthSoM/blob/main/urban\_crossroad/README.md}.

\begin{figure}[!t]
\centering
\includegraphics[width=0.99\linewidth]{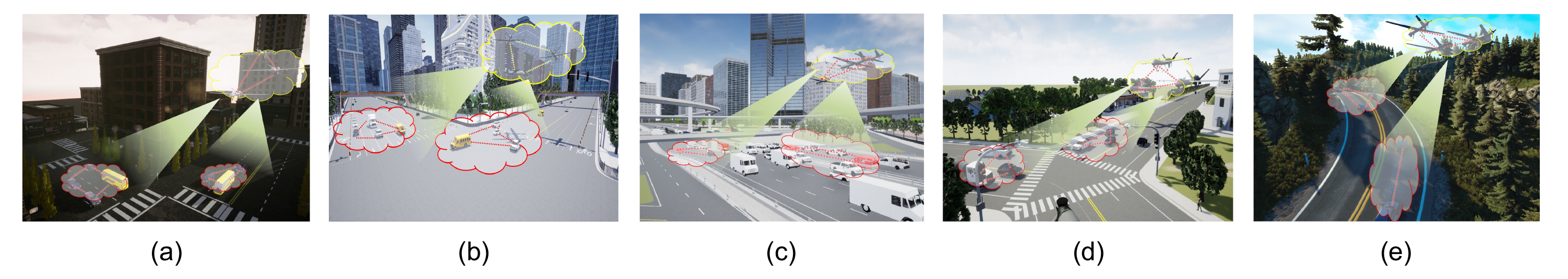}
\caption{Air-ground multi-link cooperative
scenarios in the SynthSoM dataset. (\textbf{a}) Urban crossroad scenario. (\textbf{b}) Urban wide lane scenario. (\textbf{c}) Urban overpass scenario. (\textbf{d}) Suburban fork scenario. (\textbf{e}) Mountain road scenario.}
\label{fig:scenario}
\end{figure}

To achieve the precise alignment, the high-fidelity scenarios in AirSim, WaveFarer, and Wireless InSite are accurately aligned. The size/coordinate of each object model in each software at the initial snapshot is the same.  Specifically, when exporting scenario models from UE4 to WaveFarer and Wireless InSite, the STL format is utilized for export, ensuring no rotation, scaling, or origin offset. By further aligning the coordinate systems, verifying a $1:1$ scale, and unifying the world origin, the strict consistency with the scenario models in UE4 in terms of geometric coordinates, scale, and position is guaranteed.
Currently, the SynthSoM dataset covers air-ground multi-link cooperative scenarios with five typical areas, i.e., urban crossroad, urban wide lane, urban overpass, suburban fork, and mountain road. For clarity, taking AirSim as an example, Figure~\ref{fig:scenario} shows the existing air-ground multi-link cooperative scenarios  in the SynthSoM dataset.

\subsection*{Comprehensive Scenario Condition Simulation}
The comprehensive simulation of scenario conditions, such as multiple weather conditions, times of the day, intelligent agent densities, frequency bands, and antenna types,  in the developed simulation platform is given below. 

In AirSim, Python is utilized to control the amount of rain and snow for the weather simulation. The amount of rain, degree of water accumulation, amount of snow, and degree of snow accumulation are represented by Rain, Roadwetness, Snow, and RoadSnow, respectively. By setting Rain, Roadwetness, Snow, and RoadSnow, sunny, rainy, and snowy days can be mimicked. To simulate rainy and snowy conditions on RGB images and depth maps, the RGB images and depth maps are
captured by the AirSim plugin based on the actual scenario weather, meaning that the RGB image is a snapshot
of the scenario instantaneous state, and the depth map is a snapshot of the scenario instantaneous state.
Nonetheless, because of  the inaccurate characterization of object properties in AirSim, e.g., material reflectivity, weather conditions solely affect RGB and depth cameras and not LiDAR devices. This limitation is also present in other software, e.g., CARLA \cite{dosovitskiy2017carla} and LGSVL \cite{rong2020lgsvl}. To overcome this drawback, we exploit a LiDAR point weather augmentation method, i.e., LISA \cite{kilic2021lidar}. 
The LISA method simulates the propagation process of laser beams in the physical world to determine the reflection positions and corresponding power intensities. Specifically, the Monte Carlo algorithm is utilized to simulate the distribution of scatterers, such as rain and snow particles in space, the Fresnel reflection is employed to depict the propagation path of light in space, and the position and intensity of the scatterers are measured based on the minimum detectable power of the sensor.
The rain rate and the snow rate in the LISA method are set to $50$~mm/hr and $10$~mm/hr, respectively \cite{kilic2021lidar}. In addition to different weather conditions,  AirSim is also capable of simulating various times of the day by adjusting the sun altitude and the intensity of skylight. The angle of the sun altitude influences the angle of light incidence, thus capturing the distinct effects of dawn, morning, and night. Furthermore, the intensity of skylight directly reflects the global environmental illumination and can be matched with different values to represent the varying effects of dawn, morning, and night. For the simulation of dawn, i.e., dim light during the day, the sun elevation angle $Sun_\mathrm{height}$ is set to $0$ with the sun axis aligned parallel to the horizon $Elevation$ also set to $0$. The sky light parameter $Sky_\mathrm{light}$ is set to $1.0$, which means base ambient illumination. For the simulation of morning, i.e., strong light during the day, the sun elevation angle $Sun_\mathrm{height}$ is set to $1.0$ with the sun axis aligned parallel to the horizon $Elevation$ set to $90^{\circ}$. The sky light parameter $Sky_\mathrm{light}$ is set to $2.0$, which means enhanced global illumination. For the simulation of night, i.e., nighttime environment, the sun elevation angle $Sun_\mathrm{height}$ is set to $-1$. According to the astronomical twilight criteria, the sun axis aligned parallel to the horizon $Elevation$ is set to $-18^{\circ}$, where the sun is below the horizon.  The sky light parameter $Sky_\mathrm{light}$ is set to $0.2$, which is the minimum ambient light. Furthermore, the starlight brightness parameter $Star_\mathrm{intensity}$ is set to $0.8$.
Finally, by placing different numbers of intelligent agents, we  simulate scenarios with high, medium, and low intelligent agent densities. 

To achieve the in-depth integration  between AirSim and WaveFarer, we utilize a single rain/snow model and import it into WaveFarer to check if its sizes match the current scenario. By further leveraging an open-sourced 3D computer graphics software, i.e., Blender\cite{Blender}, we  scale the individual rain/snow model to an appropriate size and combine it to create  a rain/snow model. Next, we import the processed rain and snow models into WaveFarer to simulate rainy and snowy days, respectively. Then, the temperature and humidity are set. For the rainy day, the temperature is set to $22.2$ degrees Celsius and the humidity is set to $100$\%. For the snowy day, the temperature is set to $-10$ degrees Celsius and the humidity is set to $20$\%.
To mimic high, medium, and low intelligent agent densities, we place different numbers of intelligent agents. For the setting of the antenna type, we design an antenna type and set the $3$~dB attenuation and cutoff beamwidth in the vertical and horizontal directions. Transceivers are configured based on the desired array type and the transmit power of mmWave radar is set. In the FMCW setting, parameters such as the start and stop frequencies of the scan, the duration of a single scan, and the number of scans per snapshot are selected.

To achieve the in-depth integration  between AirSim and Wireless InSite, we import the rain  and snow models that are constructed in WaveFarer into Wireless InSite to mimic rainy and snowy days, respectively. The temperature and humidity for rainy and snowy days in Wireless InSite are the same as those for rainy and snowy days in WaveFarer. Since the intelligent agent density has a significant impact on the channel data \cite{huang2023mixed,yang2021non}, we also consider high, medium, and low intelligent agent densities by placing different numbers of intelligent agents. Furthermore, in the Waveforms module of Wireless InSite, we can select the different waveform types and set different frequency bands, such as sub-6 GHz and mmWave. In the Antennas module of Wireless InSite, we can configure different antenna types, including single-input single-output (SISO), multiple-input multiple-output (MIMO), and massive MIMO. 

\begin{figure}[!t]
\centering
\includegraphics[width=0.99\linewidth]{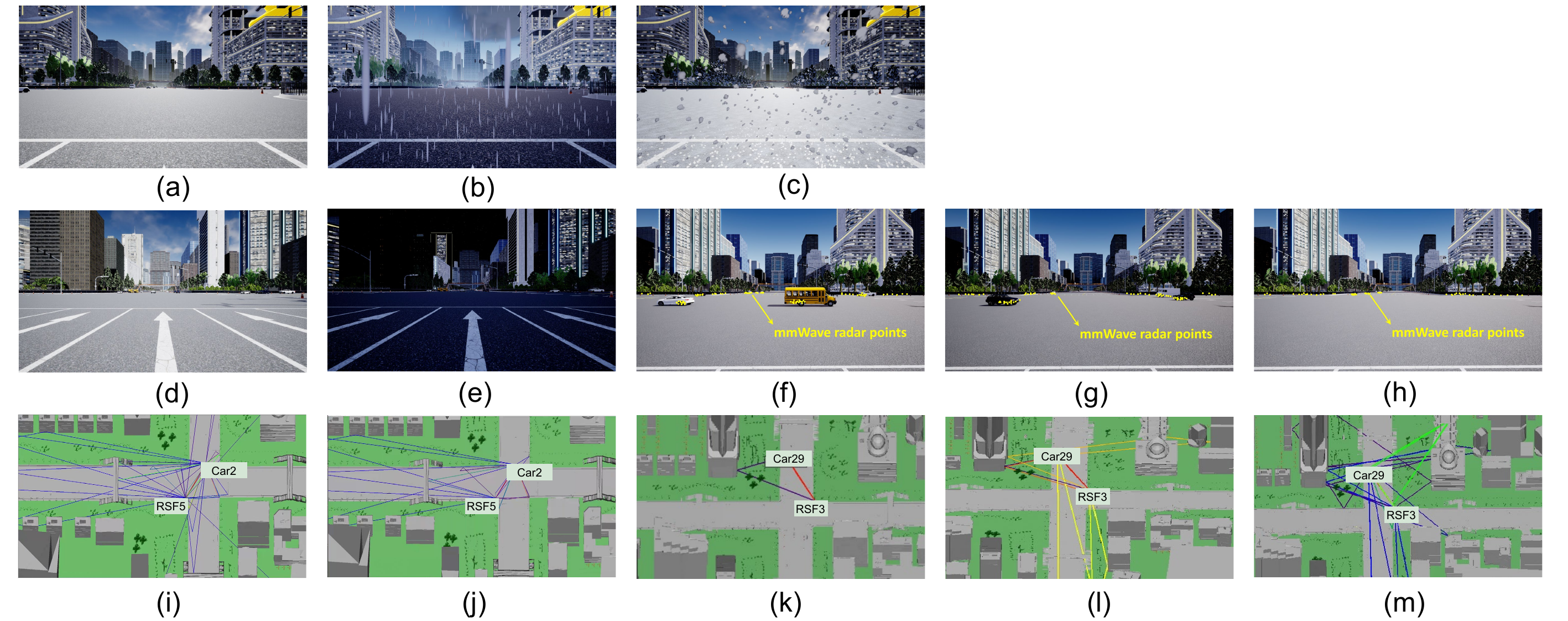}
\caption{Comprehensive conditions in the SynthSoM dataset taking the urban wide lane scenario as an example. (\textbf{a}) Sunny day. (\textbf{b}) Rainy day. (\textbf{c}) Snowy day. (\textbf{d}) Morning. (\textbf{e}) Night. (\textbf{f}) High intelligent agent density. (\textbf{g}) Medium intelligent agent density. (\textbf{h}) Low intelligent agent density. (\textbf{i}) Sub-6 GHz band. (\textbf{j}) mmWave band. (\textbf{k}) SISO condition. (\textbf{l}) MIMO condition. (\textbf{m}) Massive MIMO condition.}
\label{fig:condition}
\end{figure}

To achieve the precise alignment of AirSim, WaveFarer, and Wireless InSite, the weather conditions and intelligent agent densities in the aforementioned software are  aligned by properly setting the weather parameter and the number of intelligent agents, respectively. As a result, the multi-modal sensory data and communication data are aligned under different scenario conditions.
Currently, the SynthSoM dataset covers three weather conditions, i.e., sunny, rainy, and snowy days, two times of the day, i.e., morning and night, three types of  intelligent agent densities, i.e., high, medium, and low, two frequency bands, i.e., sub-6 GHz and mmWave, and three antenna types, i.e., SISO, MIMO, and massive MIMO. For clarity, taking the urban wide lane scenario as an example, Figure~\ref{fig:condition} illustrates comprehensive scenario conditions in the SynthSoM dataset. Note that the simultaneous incorporation and interaction  of various scenario conditions enhance the comprehensiveness of the SynthSoM dataset.

\subsection*{Dynamic Scenario Generation}
In the developed simulation platform, we set the aligned and complex trajectory of the dynamic intelligent agent in AirSim, WaveFarer, and Wireless InSite. 

To generate the realistic trajectory of the dynamic intelligent agent, we exploit an open-sourced microscopic traffic software, i.e., Simulation of Urban MObility (SUMO) \cite{lopez2018microscopic}. First, we export the road network from RoadRunner and import it into SUMO to obtain the initial geometric structure of the road network, which is further revised according to road conditions. Second, we select several positions as anchor points in the road network of RoadRunner and SUMO and measure the coordinates of these anchor points. By calculating the average of the affine transformation matrices and performing translation, rotation, and scaling, we align the coordinate systems of RoadRunner and SUMO. Third, we set traffic rules for each lane, such as straight, left turn, and right turn. Meanwhile, we install traffic lights at each intersection and differentiate lanes with solid and dashed lines. Fourth, according to  scenario requirements, we calculate simulation parameters, generate random traffic flows, and correct  abnormal traffic flows. As a result, the trajectory of the dynamic intelligent agent is generated in SUMO. 
Fifth, we import the trajectory generated in SUMO into UE4, which can be further imported into AirSim, WaveFarer, and Wireless InSite, thus achieving an in-depth integration among them. Similar to the coordinate alignment operation between RoadRunner and SUMO, we also achieve coordinate alignment between UE4 and SUMO. Finally, trajectories generated in UE4 are imported into AirSim, Wireless InSite, and WaveFarer to ensure that the three software have consistent trajectories, thus achieving a precise alignment among them.
Note that a detailed parameter setting in the third step, i.e., dynamic scenario generation, can be observed in \url{https://github.com/ZiweiHuang96/SynthSoM}. For example, for Scenario 1, i.e., Air-Ground Multi-Link Cooperative Urban Crossroad, the parameter related to intelligent agent trajectory is given and can be found in  \url{https://github.com/ZiweiHuang96/SynthSoM/blob/main/urban\_crossroad/README.md}. Therefore, for the data annotation, the label or ground truth of position and orientation parameters related to intelligent agents and RSFs is naturally contained in the synthetic SynthSoM dataset.
For vehicles, they accelerate, decelerate, and change lanes on a two-dimensional (2D) ground, thus lacking height information. Different from vehicles, UAVs fly in 3D space, possessing altitude information and exhibiting ascending and descending behaviors. Furthermore, different scenarios also present entirely different trajectories for vehicles and UAVs. Therefore, vehicles and UAVs have different trajectories, and the trajectories vary significantly across different scenarios, thus enhancing the diversity of the SynthSoM dataset.

After determining the trajectory of the dynamic intelligent agent, we generate dynamic scenarios in batches. In AirSim, we revise positions of dynamic intelligent agents and sensors via Python, and further set their 3D coordinates snapshot by snapshot. 
In WaveFarer, positions of dynamic intelligent agents and mmWave radars  are controlled via QtScript. With the help of QtScript, we import  positions and orientations of dynamic intelligent agents from external sources. By using coordinate parameters, we further fix the relative position of the mmWave radar to the corresponding dynamic intelligent agent. In Wireless InSite, based on the determined trajectory, we revise positions and orientations of both dynamic intelligent agents and antennas. Specifically, for dynamic intelligent agents, we calculate and revise coordinates based on their rotation angles using rotation matrices. As for antennas, we can directly revise their coordinates and corresponding rotation angles.

\begin{figure}[!t]
\centering
\includegraphics[width=0.99\linewidth]{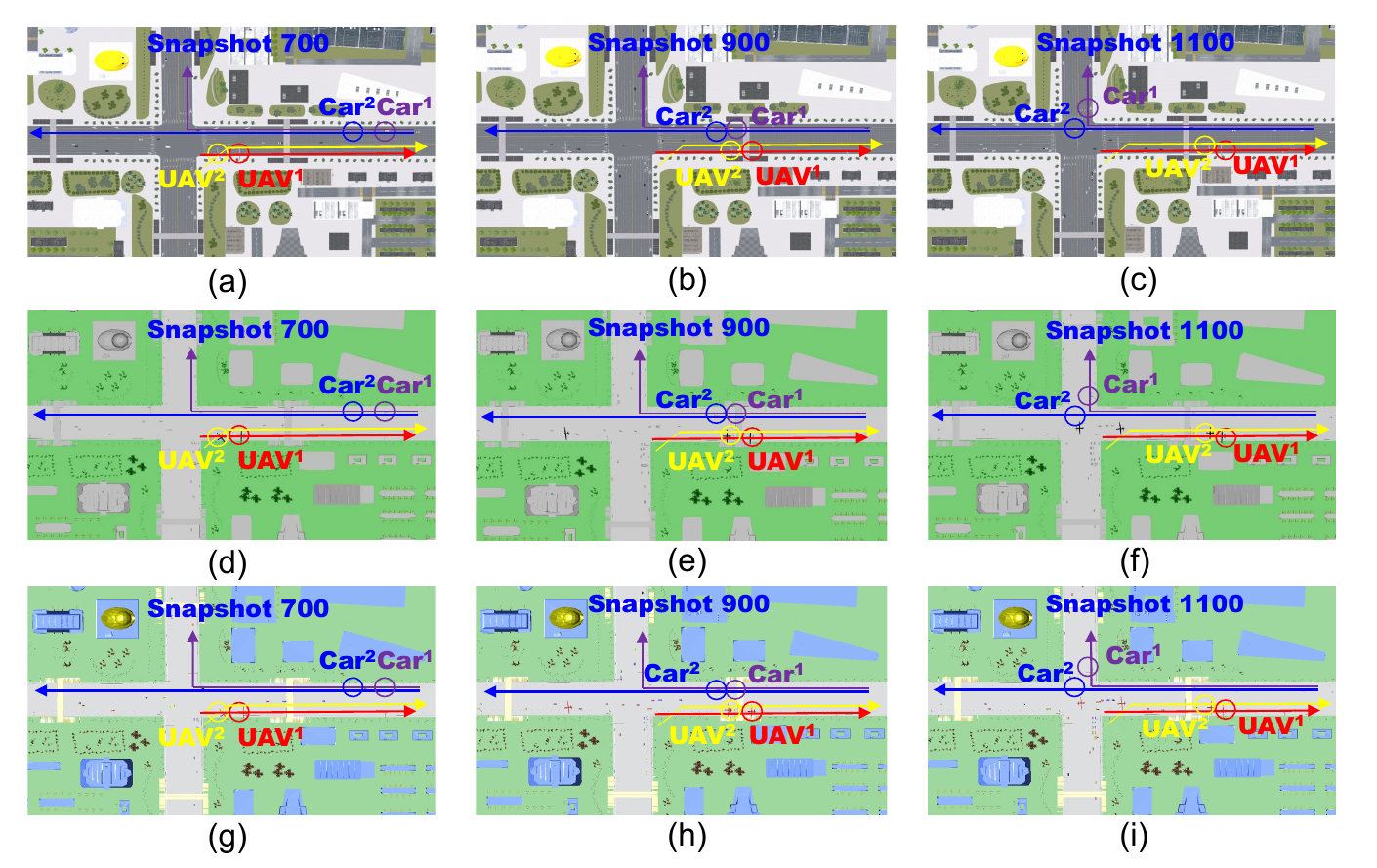}
\caption{Dynamic scenarios at Snapshot 700, Snapshot 900, and Snapshot 1100 taking the urban wide lane scenario as an example. (\textbf{a}) AirSim, Snapshot 700. (\textbf{b}) AirSim, Snapshot 900. (\textbf{c}) AirSim, Snapshot 1100. (\textbf{d}) Wireless InSite, Snapshot 700. (\textbf{e}) Wireless InSite, Snapshot 900. (\textbf{f}) Wireless InSite, Snapshot 1100. (\textbf{g}) WaveFarer, Snapshot 700. (\textbf{h}) WaveFarer, Snapshot 900. (\textbf{i}) WaveFarer, Snapshot 1100.}
\label{fig:trajectory}
\end{figure}

The SynthSoM dataset contains air-ground multi-link cooperative scenarios with complex trajectories of UAVs and vehicles. For clarity, taking the urban wide lane scenario as an example, Figure~\ref{fig:trajectory} depicts dynamic scenarios at different snapshots in the SynthSoM dataset, where the trajectories of two typical UAVs and two typical vehicles are given.

\subsection*{Data Collection and Export}
Finally, the developed simulation platform collects and exports communication data and multi-modal sensory data automatically.

In AirSim and WaveFarer, as previously mentioned,  trajectories of dynamic intelligent agents are  set. Through the predefined trajectory,  multi-modal sensors mounted on dynamic intelligent agents collect and automatically export multi-modal sensory data, including RGB images, depth maps, LiDAR point clouds, and mmWave radar point clouds.
In Wireless InSite, the generated communication scenarios are queued for simulation by a script. For the SISO condition, channel data, such as path loss and channel matrix, can be directly collected and exported. However, for the MIMO or the massive MIMO condition, Wireless InSite cannot directly  export the collected channel data and instead requires manual exportation. To overcome this limitation, we utilize Python to automatically control the  mouse to  export the collected channel data without manual exportation.
The SynthSoM dataset encompasses multiple data modalities, including RF channel large-scale and small-scale fading data, RF mmWave radar sensory data, and non-RF sensory data, e.g., RGB images, depth maps, and LiDAR point clouds. Due to the in-depth integration and precise alignment of the aforementioned three steps, i.e., high-fidelity scenario construction, comprehensive scenario condition simulation, and dynamic scenario generation, the collected communication data and multi-modal sensory data are also deeply integrated and precisely aligned.

In summary, the developed platform integrates three high-precision software, i.e., AirSim, WaveFarer, and Wireless InSite, and further achieves in-depth integration and precise alignment of them.  There are four steps involved in developing the simulation platform, including high-fidelity scenario construction (the first step), comprehensive scenario condition simulation (the second step), dynamic scenario generation (the third step), and data collection and export (the fourth step). Based on the developed platform, the SynthSoM dataset is constructed, which encompasses multiple data modalities under various air-ground multi-link cooperative  scenarios with comprehensive conditions. 
For clarity, Table~\ref{tab:example32} depicts the key algorithms and parameters utilized in the construction of the SynthSoM dataset from the perspectives of four steps.

\begin{table}[!t]
\centering
\begin{small}
\begin{tabular}{|l|l|l|l|l|}
\hline
\textbf{Software} & \textbf{First Step} & \textbf{Second Step}& \textbf{Third Step}& \textbf{Fourth Step}\\
\hline
AirSim  & \makecell[l]{Rendering pipeline of UE4 \\Emit a multitude of layer rays} & \makecell[l]{AirSim plugin \\LISA algorithm} & \multirow{5}{*}{\makecell[l]{Generate trajectories of \\ dynamic intelligent  \\agents based on  \\SUMO and UE4 }}  & \makecell[l]{Export RGB images, depth maps,\\ and LiDAR point clouds}\\
\cline{1-3} \cline{5-5} 
WaveFarer  & \makecell[l]{Ray-tracing simulation \\for modeling radar paths} & \makecell[l]{Rain model \\Snow model}&  & Export mmWave radar data  \\
\cline{1-3} \cline{5-5} 
Wireless InSite  & \makecell[l]{Ray-tracing simulation \\for modeling propagation paths} & \makecell[l]{Rain model \\Snow model}&  & Export  communication channel data\\
\hline
\multicolumn{5}{|l|}{\makecell[l]{Detailed parameter setting can be observed in \url{https://github.com/ZiweiHuang96/SynthSoM}.}
}\\
\hline
\end{tabular}
\caption{\label{tab:example32}Key algorithms and parameters utilized in the construction of the SynthSoM dataset.}
\end{small}
\end{table}


 

\section*{Data Records}
\begin{figure}[!t]
\centering
\includegraphics[width=0.8\linewidth]{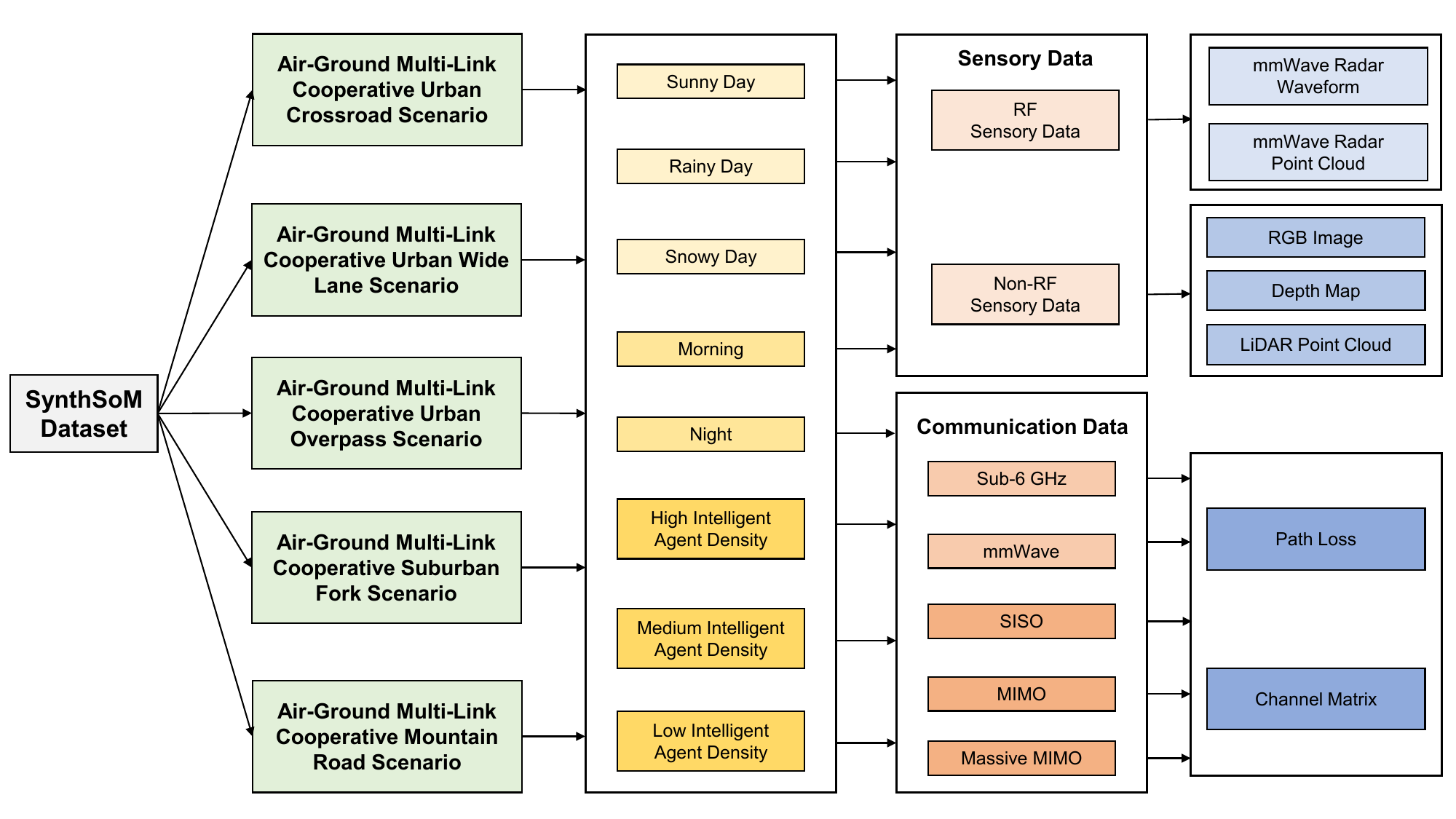}
\caption{Data directory hierarchy and content of the SynthSoM dataset.}
\label{fig:directory}
\end{figure}

The SynthSoM dataset is available at the Figshare repository\cite{Figshare_new}. To utilize the SynthSoM dataset for communication and multi-modal sensing tasks, we divide the SynthSoM dataset into five folders according to different scenarios. The scenario folder contains different conditions, and further contains  different data modalities, i.e., RF sensing, non-RF sensing, and RF communications. The data directory hierarchy and content of the SynthSoM dataset are shown in Fig.~\ref{fig:directory}. Due to the significantly massive data amount of the SynthSoM dataset, we utilize an efficient compression way and upload the SynthSoM dataset to the Figshare repository\cite{Figshare_new} in 7-zip compression format. Overall, the constructed SynthSoM dataset is fully open-sourced and has a size of 743.79 GB.

Multi-modal sensory data contains RF sensory data, i.e., mmWave radar waveforms and point clouds, and non-RF sensory data, i.e., RGB images, depth maps, and LiDAR point clouds. For RF sensory data, the mmWave radar waveform is stored in csv file format and the mmWave radar point cloud is stored in mat file format. For non-RF sensory data, the RGB image is stored in png file format and the LiDAR point cloud is  stored in txt file format. Meanwhile, the depth map includes the DepthPerspective file and the DepthPlanar file, which are stored in pfm and png formats, respectively. To simultaneously consider channel large-scale and small-scale fading characteristics, communication data contains the path loss and channel matrix. The path loss  is stored in txt file format and the channel matrix  is stored in mat file format. 

Overall, the SynthSoM dataset contains 140K sets of channel matrices, 18K sets of path loss, 136K sets of mmWave radar waveforms with 38K radar point clouds, 145K RGB images, 290K depth maps, and 79K sets of LiDAR point clouds. The open-sourced SynthSoM dataset can provide consistent data for SoM-related algorithm cross comparison, model calibration, and baseline implementation.


\section*{Technical Validation}
We validate the accuracy of the SynthSoM dataset in two ways. In the first subsection, from the statistics-based qualitative inspection perspective, we compare the visualization and statistical distribution of the synthetic data in the SynthSoM dataset to the measurement data in the real world. In the second subsection, from the ML-based evaluation metric perspective, we conduct the ML experiment to investigate whether the performance on the synthetic data in the SynthSoM dataset can be transferred to the real world. 

\subsection*{Statistics-Based Qualitative Inspection}
For the communication data in the SynthSoM dataset, the visualization and statistical properties of channels under different conditions are obtained. For the multi-modal sensory data in the SynthSoM dataset, mmWave radar data, RGB images, depth maps, and LiDAR point clouds are visualized. The obtained and visualized result and phenomenon from the SynthSoM dataset are consistent with the real world. Therefore, the SynthSoM dataset can be validated via the statistics-based qualitative inspection.

\begin{figure}[!t]
\centering
\includegraphics[width=0.99\linewidth]{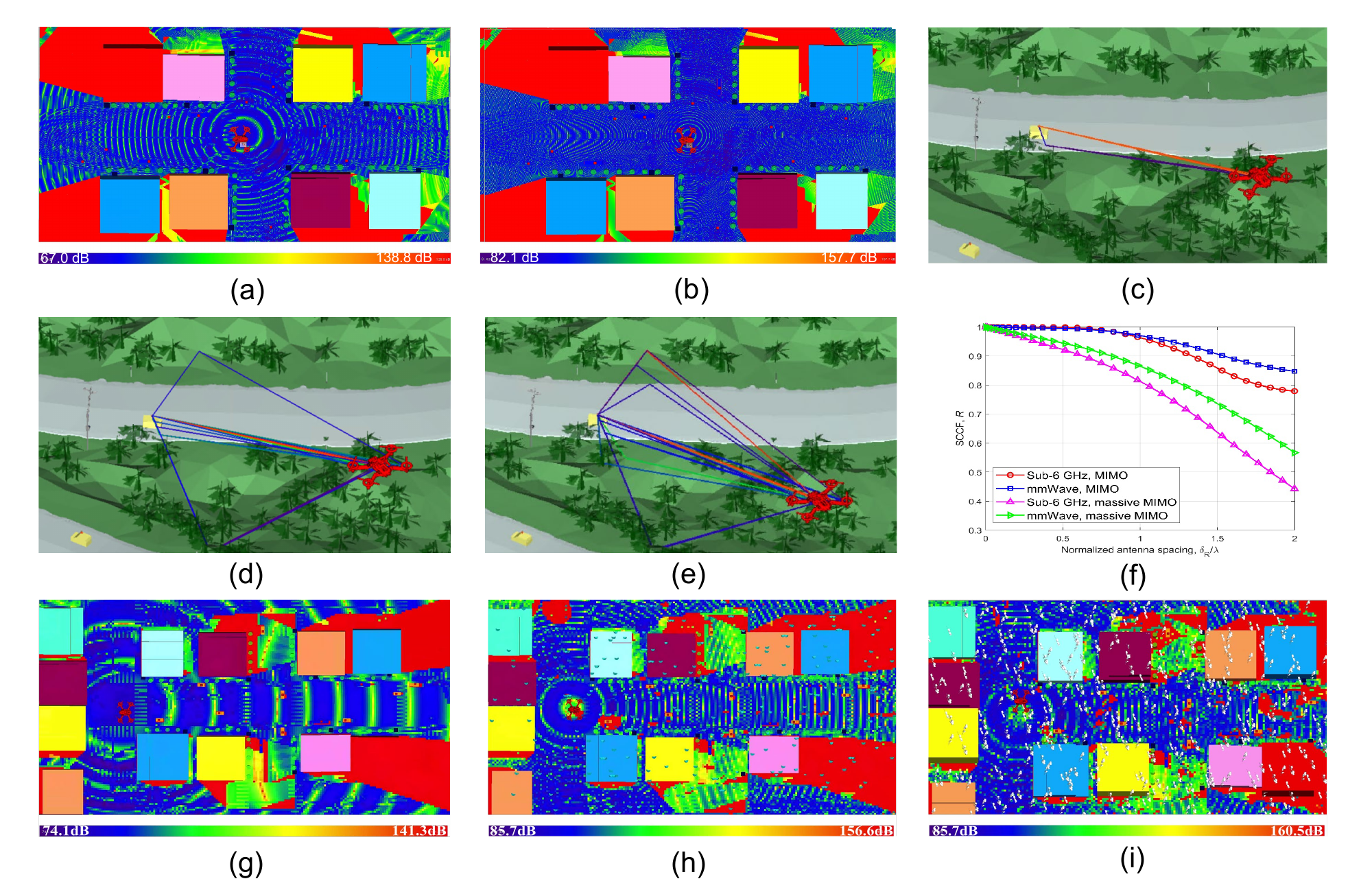}
\caption{Visualization and statistical properties of RF channels under different conditions. (\textbf{a}) Path loss heatmap in urban crossroad scenario with sub-6 GHz frequency band. (\textbf{b}) Path loss heatmap in urban crossroad scenario with mmWave frequency band. (\textbf{c}) Propagation paths in mountain road scenario under SISO. (\textbf{d}) Propagation paths in mountain road scenario under MIMO. (\textbf{e}) Propagation paths in mountain road scenario under massive MIMO. (\textbf{f}) SCCFs in mountain road scenario with sub-6 GHz and mmWave frequency bands under MIMO and massive MIMO conditions. (\textbf{g}) Path loss heatmap in urban crossroad scenario on sunny days. (\textbf{h}) Path loss heatmap in urban crossroad scenario on rainy days. (\textbf{i}) Path loss heatmap in urban crossroad scenario on snowy days.}
\label{fig:communications}
\end{figure}

Figure~\ref{fig:communications} depicts the visualization and statistical properties of RF channels under different conditions. Figures~\ref{fig:communications}(a) and (b) give path loss heatmaps in the urban crossroad scenario with sub-6 GHz and mmWave frequency bands, respectively. Since path loss is proportional to frequency, the path loss under  mmWave frequency band, ranging from $82.1$ dBm to $157.7$ dBm, is higher than that with sub-6 GHz frequency band, ranging from $67.0$ dBm to $138.8$ dBm. This phenomenon is consistent with the measurement result \cite{de2021radio}. Figures~\ref{fig:communications}(c)--(e) show propagation paths in the mountain road scenario under SISO, MIMO, and massive MIMO, respectively. With an increase in the number of antennas, the number of propagation paths increases, leading to a larger spatial diversity gain. This phenomenon is in agreement with the measurement result \cite{yan2015measurement}. Figure~\ref{fig:communications}(f) compares space cross-correlation functions (SCCFs) in the mountain road scenario with sub-6 GHz and mmWave frequency bands under MIMO and massive MIMO conditions. The SCCF with sub-6 GHz frequency band is lower than that with mmWave frequency band, which is the consistent with the measurement result \cite{miao2023sub}. This is because that, compared to mmWave frequency band, there are more propagation paths in sub-6 GHz frequency band, resulting in larger channel spatial diversity and lower spatial correlation. Similarly, the SCCF under massive MIMO condition with more propagation paths is lower than that under MIMO condition. Figures~\ref{fig:communications}(g)--(i) demonstrate path loss heatmaps in the urban crossroad scenario  on sunny, rainy, and sunny days, respectively. In comparison with sunny days, path loss is much smaller on rainy and snowy days attributed to the obvious rain and snow attenuation at the mmWave frequency band. This is consistent with the measurement result \cite{PLWeather1,PLWeather2}.

\begin{figure}[!t]
\centering
\includegraphics[width=0.99\linewidth]{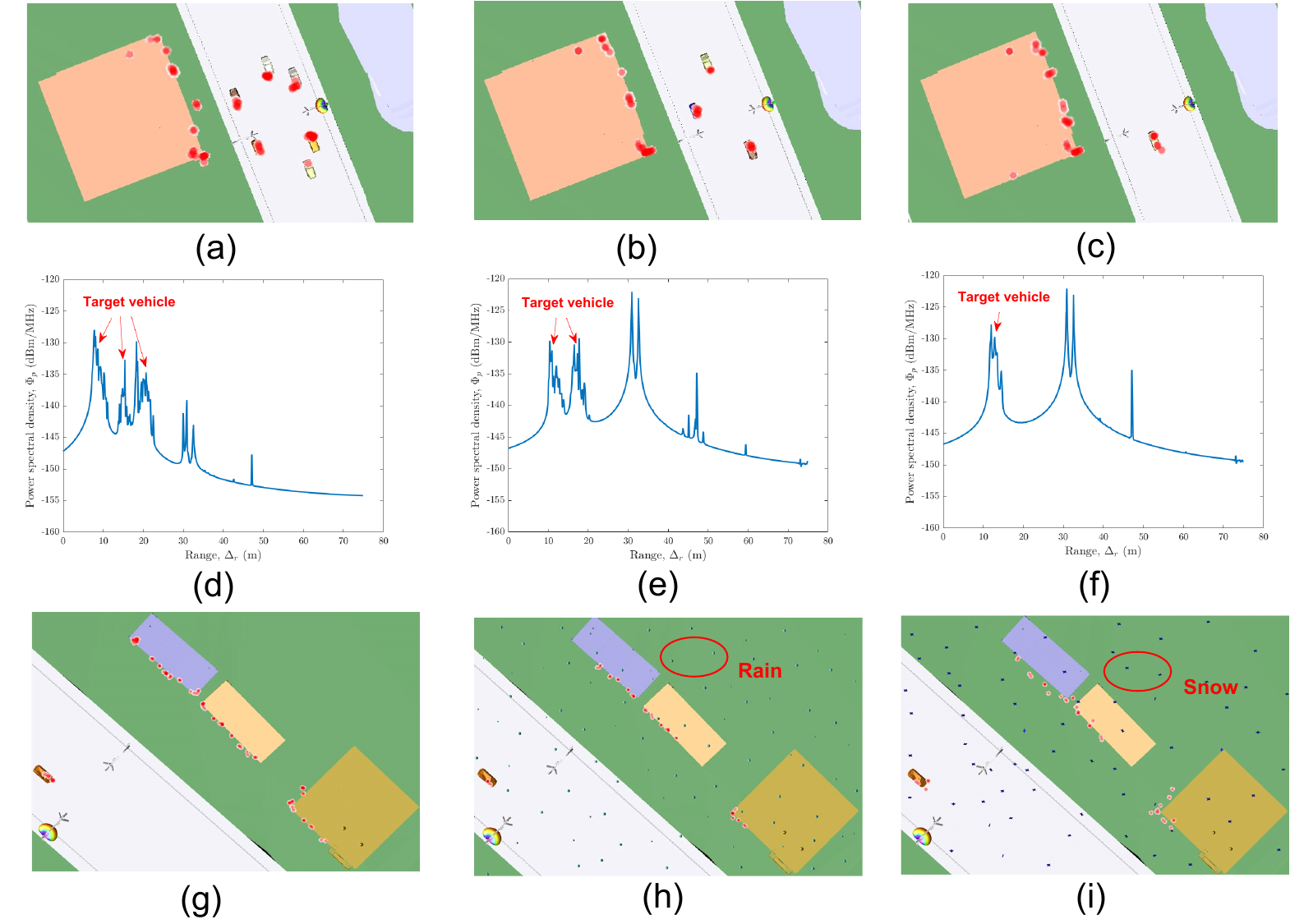}
\caption{Visualization of RF sensory data under different intelligent agent densities. (\textbf{a}) mmWave radar point cloud under high intelligent agent density. (\textbf{b}) mmWave radar point cloud under medium intelligent agent density. (\textbf{c}) mmWave radar point cloud under low intelligent agent density. (\textbf{d}) Range-power map under high intelligent agent density. (\textbf{e}) Range-power map under medium intelligent agent density. (\textbf{f}) Range-power map under low intelligent agent density. (\textbf{g}) mmWave radar point cloud on sunny days. (\textbf{h}) mmWave radar point cloud on rainy days. (\textbf{i}) mmWave radar point cloud on snowy days.}
\label{fig:RF sensing}
\end{figure}

Figure~\ref{fig:RF sensing} gives the visualization of RF sensory data under different intelligent agent densities.
For clarity, the suburban fork scenario is taken as an example for analysis. Figures~\ref{fig:RF sensing}(a)--(c) show mmWave radar point clouds under high, medium, and low intelligent agent densities at an intersection, respectively. Compared to the low and medium intelligent agent densities, the intelligent agent number within the  FoV of mmWave radar is greater in the high intelligent agent density, leading to a higher number of detected mmWave radar points. 
Figures~\ref{fig:RF sensing}(d)--(f) correspond to the range-power maps of Figs.~\ref{fig:RF sensing}(a)--(c), respectively. As the intelligent agent number detected by the mmWave radar increases, the number of power spectrum density peaks increases, which is consistent with the measurement result \cite{gao2019experiments}. Furthermore, the power spectrum density peak corresponding to the low-rise building approximately $32$ m away on the opposite side of the road decreases with the increase in the intelligent agent density. This is due to the increased blockage caused by the intelligent agent in the middle of the road. Figures~\ref{fig:RF sensing}(g)--(i) depict mmWave radar point clouds on sunny, rainy, and snowy days, respectively. 
The scattering and obstruction of electromagnetic waves are minimal, leading to precise object detection and a longer detection range on sunny days. However, the scattering and absorption of mmWave radar signals by raindrops on rainy days reduce the maximum detection range of the mmWave radar \cite{WFweather11}. Also, the scattering and absorption of mmWave radar signals by snowflakes are more pronounced on snowy days, thus further reducing the maximum detection range and the accuracy of object detection \cite{WFweather22}.

\begin{figure}[!t]
\centering
\includegraphics[width=0.99\linewidth]{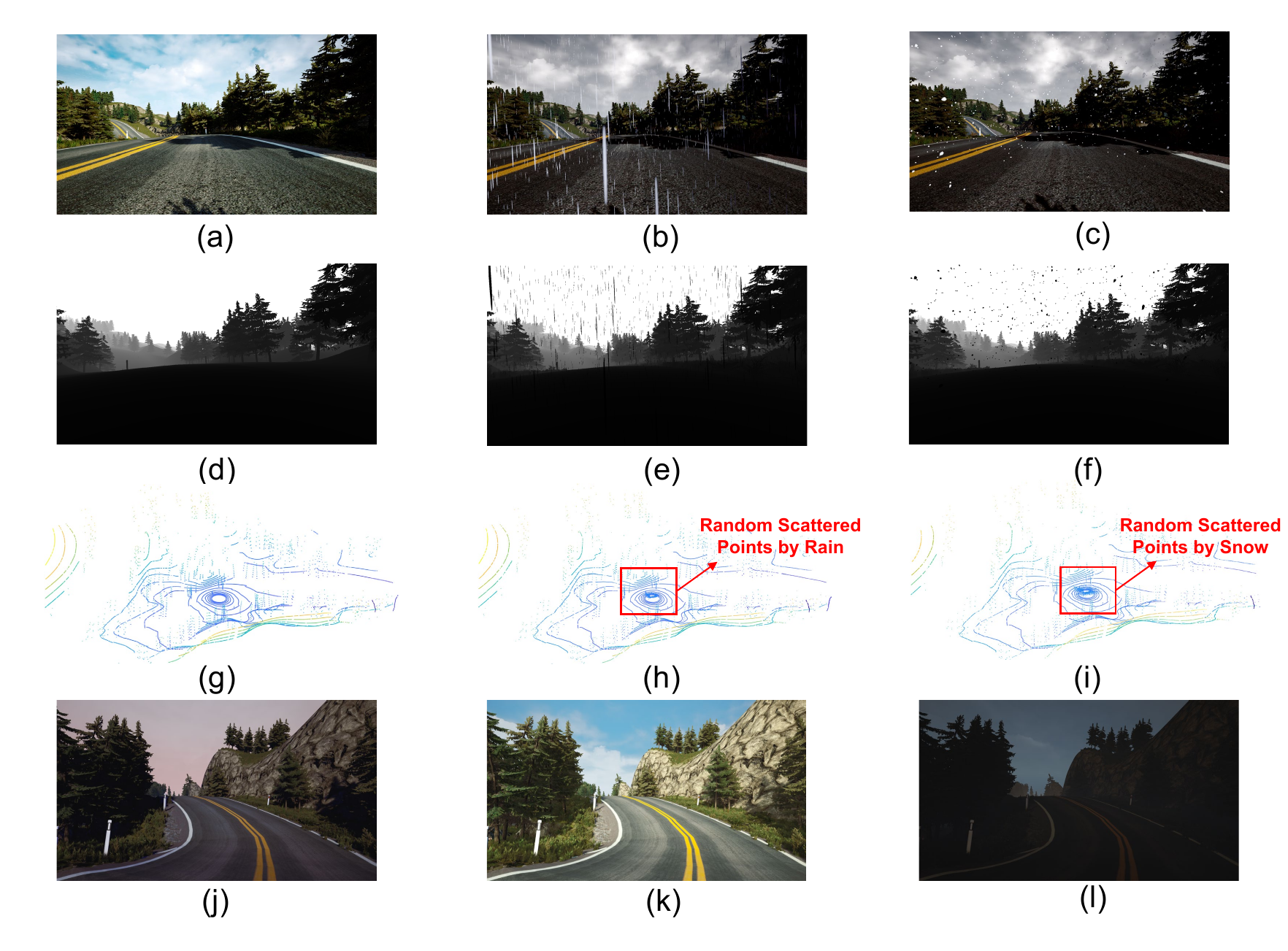}
\caption{Visualization of non-RF sensory data under different conditions. (\textbf{a}) RGB image on sunny days. (\textbf{b}) RGB image  on rainy days. (\textbf{c}) RGB image  on snowy days. (\textbf{d}) Depth map on sunny days. (\textbf{e}) Depth map  on rainy days. (\textbf{f}) Depth map  on snowy days. (\textbf{g}) LiDAR point cloud on sunny days. (\textbf{h}) LiDAR point cloud on rainy days. (\textbf{i}) LiDAR point cloud on snowy days. (\textbf{j}) RGB image  at dawn. (\textbf{k}) RGB image  in the morning. (\textbf{l}) RGB image  at night.}
\label{fig:non-RF sensing}
\end{figure}

Figure~\ref{fig:non-RF sensing} illustrates the visualization of non-RF sensory data in the mountain road scenario under different times of the day and weather conditions. Figures~\ref{fig:non-RF sensing}(a)--(c) show RGB images on sunny, rainy, and snowy days, respectively. Figures~\ref{fig:non-RF sensing}(d)--(f) show depth maps on sunny, rainy, and snowy days, respectively. On rainy and snowy days, there are noise data in the RGB image and the depth map, which is consistent with the measurement result \cite{zhang2023perception}. Figures~\ref{fig:non-RF sensing}(g)--(i) show LiDAR point clouds  on sunny, rainy, and snowy days, respectively. Also, the rainy and snowy weather conditions  have a significant impact on LiDAR point clouds. Specifically, the laser attenuation resulted from random scattering medium leads to the increase in range uncertainty and loss of points far away from the LiDAR device. The huge backscattered power from random droplets further leads to the presence of random scattered points near the LiDAR device.
The aforementioned phenomenon related to the LiDAR point cloud is also in agreement with the measurement result \cite{kilic2021lidar}. Finally, Figures~\ref{fig:non-RF sensing}(j)--(l) show RGB images at dawn, morning, and night, respectively. It is certain that the visibility of RGB image in the morning is higher than that of RGB images at dawn/night.

\subsection*{ML-Based Evaluation Metrics: Performance Transferability}

Aiming at investigating if the performance on the SynthSoM dataset can be transferred to the real world, we carry out the ML-based evaluation from the perspective of performance transferability, i.e., train on synthetic, test on real (TSTR), and train on real, test on real (TRTR) \cite{hartmann2018eeg}. The motivation is straightforward, which is to train the ML model on the synthetic data in the SynthSoM dataset and evaluate its performance on the measurement data in the real world. To provide a baseline, we also train and test the ML model exclusively on the measurement data in the real world.

\begin{figure}[!t]
\centering
\includegraphics[width=0.8\linewidth]{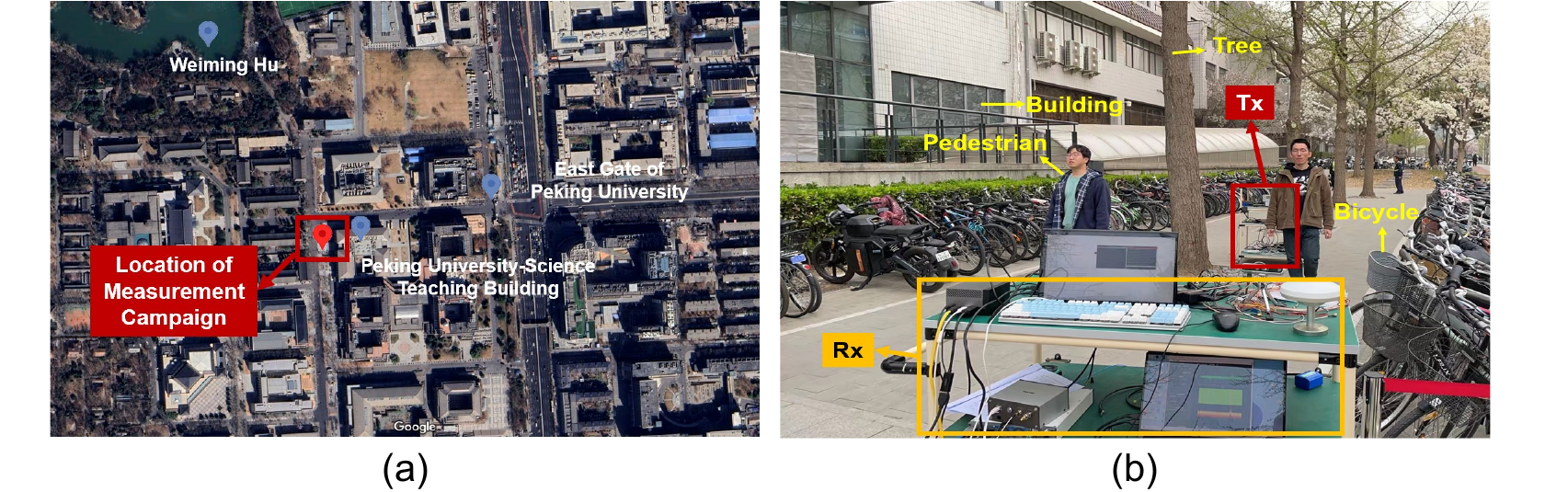}
\caption{Measurement campaign at the campus of Peking University. (\textbf{a}) Location of the measurement campaign. (\textbf{b}) Propagation environment of the measurement.}
\label{fig:PL_scenario}
\end{figure}

Collecting  communication and multi-modal sensory data in the real world requires various expensive measurement devices. These measurement devices further require spatial-temporal calibration under dynamic scenarios, which is of huge difficulty. As a result, it is expensive and difficult to collect  communication and multi-modal sensory data under dynamic scenarios in the real world. To overcome this challenge, we  conduct a measurement campaign at the campus of Peking University to collect communication and multi-modal sensory data under dynamic scenarios, as shown in Fig.~\ref{fig:PL_scenario}(a). Although the measurement scenario is not specific to the vehicular scenario, the transceiver is mobile with dynamic pedestrians and static trees/buildings in the vicinity, as shown in Fig.~\ref{fig:PL_scenario}(b). Therefore, we try our best to approximate the measurement scenario to a  dynamic vehicular scenario. Strictly speaking, the velocity of the transceiver in the vehicular scenario is higher than that in the conducted measurement campaign. Nonetheless, we are not principally interested in predicting the path loss outcome. Conversely, the ML experiment aims to demonstrate that the performance on the synthetic data can be transferred to the real world, which is similar to the performance transferability \cite{jensen2023synthetic}.  In the measurement campaign, we simultaneously utilize the sensor, communication equipment, and control device. The sensor includes an RGB-D camera, which collects RGB images with 1280$\times$720 resolution and depth maps with 848$\times$480 resolution. The communication equipment, which functions as the main RF transceiver, contains a pair of universal software radio peripheral (USRP) X410 devices and a pair of frequency converters connected to the USRP, a phased-array antenna connected to the frequency converter in Tx and a horn antenna connected to the frequency converters in Rx. As a result, the path loss between Tx and Rx can be collected, where the carrier frequency is $28$~GHz with $200$~MHz communication bandwidth. The control device consists of a pair of x86 industrial personal computers (IPCs) for controlling the transmission/reception of signals and storage of communication data, i.e., path loss, along with an ARM IPC for controlling the collection and storage of sensory data, i.e., RGB images and depth maps. By utilizing the sensor, communication equipment, and control device, we collect $882$ snapshots of aligned RGB images, depth maps, and path loss in the real world. 

To correspondingly generate and collect communication and multi-modal sensory synthetic data, we utilize the developed simulation platform. The scenario in the simulation platform is consistent with the measurement campaign, i.e., the campus of Peking University. Specifically, we import terrain information from Google Maps and construct aligned campus scenario in the simulation platform.  
Consistent with the measurement campaign, we incorporate various objects between the transceiver, including static trees/buildings and dynamic pedestrians. For a fair comparison, we set the key parameters of the simulation testing, such as the distance between the transceiver, the snapshot of collected data, the number of pedestrians, carrier frequency, communication bandwidth, etc., to be the same as those of the measurement campaign.

\begin{table}[!t]
\centering
\begin{tabular}{|l|l|}
\hline
\textbf{Hyperparameter} & \textbf{Setting}\\
\hline
Optimizer  & Adam \\
\hline
Batch Size  & 2 \\
\hline
Number of Hidden Layers  & 3 \\
\hline
Activation Function  & ReLU \\
\hline
Epoch  & 40 \\
\hline
Learning Rate  & 0.01 \\
\hline
Loss Function  & MESLoss \\
\hline
\end{tabular}
\caption{\label{tab:example2}Hyperparameters of MLPs for both TSTR and TRTR.}
\end{table}

\begin{figure}[!t]
\centering
\includegraphics[width=0.8\linewidth]{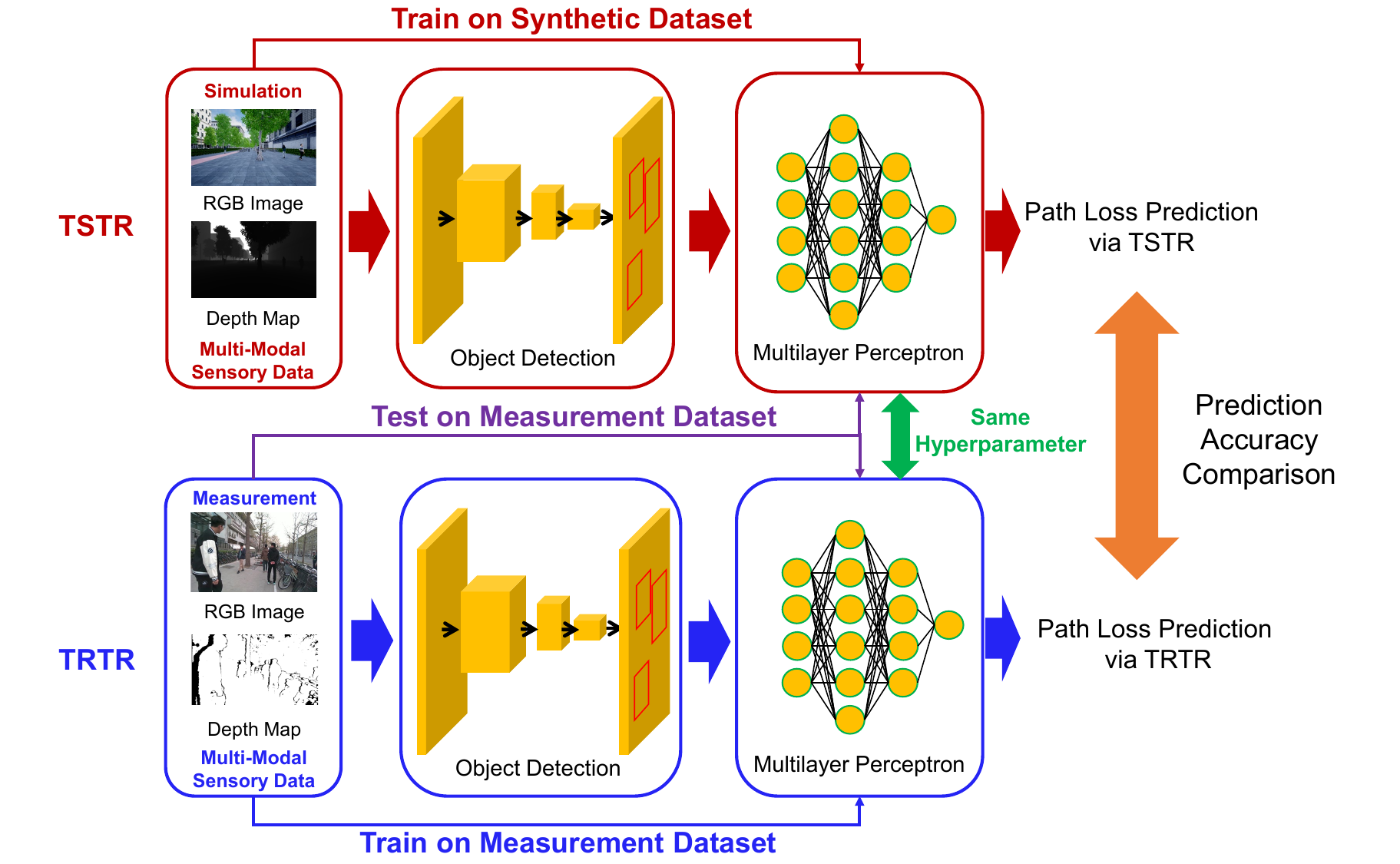}
\caption{ML experiments of TSTR and TRTR.}
\label{fig:test_PL}
\end{figure}

Based on the collected RGB image, depth map, and path loss in the real world and the developed simulation platform, we conduct the ML experiment for performance transferability testing, i.e., TSTR and TRTR,  as shown in Fig.~\ref{fig:test_PL}. For TSTR, based on an object detection network, i.e., You Only Look Once (YOLO)-v5, we utilize simulated RGB images and depth maps to extract the environmental feature and path loss, which are utilized for training and validation via multilayer perceptron (MLP). Furthermore, we evaluate the path loss prediction result of the MLP trained on the synthetic data using the testing set from the measurement data. For TRTR, we exploit the measurement data for training, validation, and testing to obtain the path loss prediction result via YOLO-v5 and MLP. For a fair comparison, the hyperparameters of MLPs for both TSTR and TRTR are consistent, as shown in Table~\ref{tab:example2}. Meanwhile, the dataset is divided into the training set, validation set, and testing set in the proportion of $3 : 1 : 1$ for both TSTR and TRTR. Therefore, the training set, validation set, and testing set contain 530 snapshots, 176 snapshots, and 176 snapshots of data, respectively. 

\begin{figure*}[!t]
\centering
\includegraphics[width=0.8\linewidth]{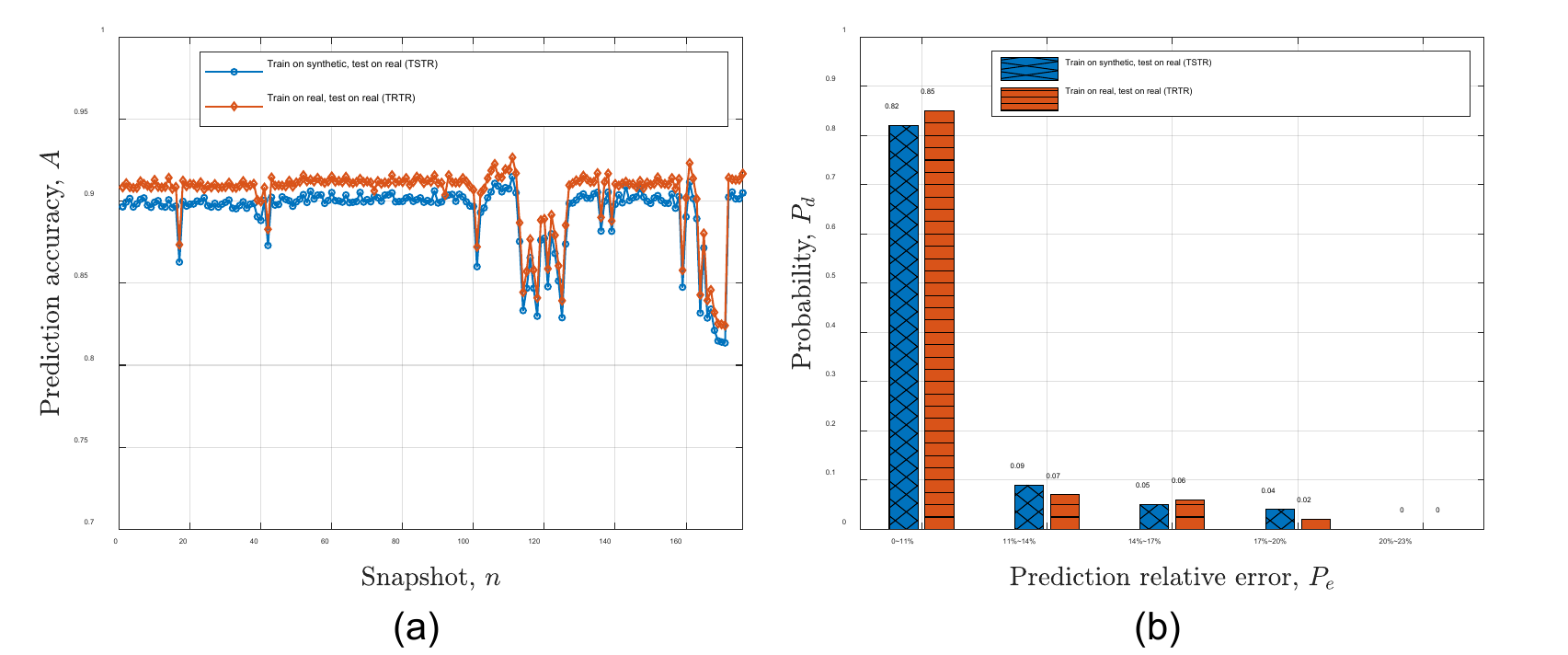}
\caption{Path loss prediction results of TSTR and TRTR. (\textbf{a}) Prediction accuracy at all snapshots of data in the testing set. (\textbf{b}) Probability of prediction relative error at all snapshots of data in the testing set.}
\label{fig:test_PL_result}
\end{figure*}

The accuracy of path loss prediction can be computed by $ A = \frac{\left|PL_\mathrm{pr}-PL_\mathrm{gt} \right|}{P_\mathrm{gt}}$, where $PL_\mathrm{pr}$ is the prediction result and $PL_\mathrm{gt}$ is the ground truth. Overall, the accuracies of path loss prediction via TSTR and TRTR are $89.28\%$ and $90.35\%$, respectively. Figure~\ref{fig:test_PL_result}(a) depicts the prediction accuracies at all snapshots, i.e., 176 snapshots, of data in the testing set via TSTR and TRTR. Figure~\ref{fig:test_PL_result}(b) further presents the probabilities of prediction relative error at all snapshots of data in the testing set via TSTR and TRTR. It can be seen from Figs.~\ref{fig:test_PL_result}(a) and (b) that path loss prediction accuracies are highly similar between synthetic data in the SynthSoM dataset and measurement data.
Also, the TSTR performs worse than the TRTR, which is consistent with the result \cite{jensen2023synthetic}. To be more specific, we observe a gap of less than 10\% in accuracy between synthetic and measurement datasets, which is acceptable for a synthetic dataset, e.g., Table~6 \cite{lee2021ctgan} and Tables~4, 5 \cite{yuan2023synthetic}. As a consequence, the comparison indicates that the performance on the synthetic data in the SynthSoM dataset can be transferred to the real world. The corresponding code  is publicly available on GitHub \url{https://github.com/ZiweiHuang96/SynthSoM/tree/main/code}.

\section*{Usage Notes}
With the help of in-depth integration and precise alignment of AirSim, WaveFarer, and Wireless InSite, the SynthSoM dataset is accurately constructed. The technical validation further indicates that the SynthSoM dataset is consistent with measurement data via statistics-based qualitative inspection and the performance on the SynthSoM dataset can be transferred to the real world via ML-based evaluation metrics. As a result, the SynthSoM dataset can be utilized to investigate a number of open problems for SoM research in AIoT applications. In the SynthSoM dataset, all data are published in the txt/mat/png/pfm/csv file format. For convenience, researchers can access the SynthSoM dataset entirely without any further permission.  Since the SynthSoM dataset is significantly large, we divide it into five folders according to different scenarios. In this case, researchers can utilize the SynthSoM dataset based on their scenarios of interest.   Therefore, by providing communication and multi-modal sensory data, the SynthSoM dataset can facilitate the SoM-related algorithm cross comparison, model calibration, and baseline implementation.




\section*{Code availability}
A git repository is publicly available on GitHub \url{https://github.com/ZiweiHuang96/SynthSoM/tree/main/code}. In this repository, based on the SynthSoM dataset, several Python and MATLAB scripts for benchmarking, visualization, and data pre-processing are available.



\section*{Acknowledgements} 

This work was supported in part by the National Natural Science Foundation of China under Grant 62125101, Grant 62341101, and Grant 62371273; in part by the Taishan Scholars Program under Grant tsqn202312307; in part by the Young Elite Scientists Sponsorship Program by CAST under Grant YESS20230372; in part by the Shandong Natural Science Foundation under Grant ZR2023YQ058; in part by the New Cornerstone Science Foundation through the XPLORER PRIZE; in part by the Xiaomi Young Talents Program; in part by the open research fund of National Mobile Communications Research Laboratory, Southeast University (No. 2025D04); in part by  the Beijing Natural Science Foundation under Grant 4254067; in part by  the China National Postdoctoral Program for Innovative Talents under Grant BX20240007; and in part by  the China Postdoctoral Science Foundation under Grant 2024M760111. The authors would like to thank Mengyuan Lu, Xujing Jia, and Ye Chen for their help in the collection of communication data via Wireless InSite and Weibo Wen, Xuanyu Liu, Junliang Lu,  and Beinan Su for their help in the collection of sensory data via AirSim and WaveFarer.

\section*{Author contributions}
X.C., Z.H., and L.B. conceived the experiments, Y.Y., M.S., Z.H., R.Z., and S.L. conducted the experiments, X.C., Z.H., and L.B. analysed the results. All authors discussed the results and commented on the manuscript. 

\section*{Competing interests}

The authors declare no competing interests.

\section*{Additional information}

\textbf{Correspondence} and requests for materials should be addressed to X.C. or L.B.

\end{document}